\documentclass[fleqn,usenatbib]{mnras}

\usepackage{newtxtext,newtxmath}
\usepackage[T1]{fontenc}
\usepackage{ae,aecompl}

\usepackage{graphicx}
\usepackage{gensymb}
\usepackage{longtable}
\usepackage{amsmath}
\usepackage{graphbox}
\usepackage{amssymb}

\title[G107.0+9.0: A New Galactic SNR]{G107.0+9.0: A New Large Optically Bright, Radio and X-Ray Faint Galactic Supernova Remnant in Cepheus}

\author[Fesen et al.]{Robert A. Fesen,$^{1}$\thanks{E-mail: robert.fesen@dartmouth.edu}
Kathryn E. Weil$^{1}$,
John C. Raymond$^{2}$,
Laurent Huet$^{3}$,
Martin Rusterholz$^{4}$,
\newauthor{Dennis di Cicco$^{5}$,}
David Mittelman$^{5}$,
Sean Walker$^{5}$,
Marcel Drechsler$^{6}$ and
Sheldon Faworski$^{7}$
\\
$^{1}$ 6127 Wilder Lab, Department of Physics and Astronomy, Dartmouth
                 College, Hanover, NH 03755, USA\\
$^{2}$ Harvard-Smithsonian Center for Astrophysics, 60 Garden Street, Cambridge, MA 02138, USA\\
$^{3}$ Pond Observatory, 28120 Illiers-Combray, France\\
$^{4}$ Cxielo Observatory, 26510 Vauclause, France\\
$^{5}$ MDW Sky Survey, New Mexico Skies Observatory, Mayhill, NM 88339, USA\\
$^{6}$ Baerenstein Observatory, Feldstrasse 17, D-09471 Baerenstein, Germany\\
$^{7}$ MASIL Observatory Midwest, 5888 S. Tower Rd., Elizabeth, IL 61028, USA\\
}

\date{Accepted XXX. Received YYY; in original form ZZZ}

\pubyear{2020}

\begin{document}
\label{firstpage}
\pagerange{\pageref{firstpage}--\pageref{lastpage}}
\maketitle

\begin{abstract}
Wide-field H$\alpha$ images of the Galactic plane have revealed a new supernova remnant (SNR) nearly three degrees in diameter centred at $l$ = 107.0, $b$ = +9.0. Deep  and higher resolution H$\alpha$ and [O~III] 5007 \AA \ images show dozens of  H$\alpha$ filaments along the remnant's northern, western, and southwestern limbs, but few  [O~III] bright filaments. The nebula is well detected in the H$\alpha$ Virginia Tech Spectral-Line Survey images, with many of its brighter filaments even visible on Digital Sky Survey images. Low-dispersion spectra of several filaments show either Balmer dominated, non-radiative filaments or the more common SNR radiative filaments with [S~II]/H$\alpha$ ratios above 0.5, consistent with shock-heated line emission. Emission line ratios suggest shock velocities ranging from $\leq70$ km s$^{-1}$  along its western limb to $\simeq100$ km s$^{-1}$ along its northwestern boundary. While no associated X-ray emission is seen in ROSAT images, faint 1420 MHz radio emission appears coincident with its western and northern limbs. Based on an analysis of the remnant's spatially resolved H$\alpha$ and [O~III] emissions, we estimate the remnant's distance at $\sim1.5 - 2.0$  kpc implying a physically large (dia. = $75- 100$ pc) and old ($90 - 110 \times 10^{3}$ yr) SNR in its post-Sedov radiative phase of evolution expanding into a low density
interstellar medium ($n_{\rm 0} = 0.05 - 0.2$ cm$^{-3}$) and lying some $250 - 300$ pc above the Galactic plane.

\end{abstract}
\bigskip


\section{Introduction}

Of the nearly 300 Galactic supernova remnants (SNRs) currently identified, the
overwhelming majority were discovered through radio observations based on the
characteristic nonthermal radio emission associated with shocked gas
\citep{Downes1971,Chevalier1977,Green1984,Green2004,Green2019}. Although less than half
of Galactic SNRs show any appreciable optical emission, optical studies of SNRs
are useful for measuring shock and expansion velocities, ISM abundances, and
the kinematics of metal-rich ejecta in the youngest remnants.  In some cases, a
remnant's optical emission can also help define its overall size and
morphology. This is especially true for some radio weak SNRs which nonetheless exhibit
significant optical emission
\citep{Boumis2009,Stupar2012,Neustadt2017,Stupar2018,How2018,Fesen2019}.

\begin{figure*}
\begin{centering}
\includegraphics[width=\textwidth]{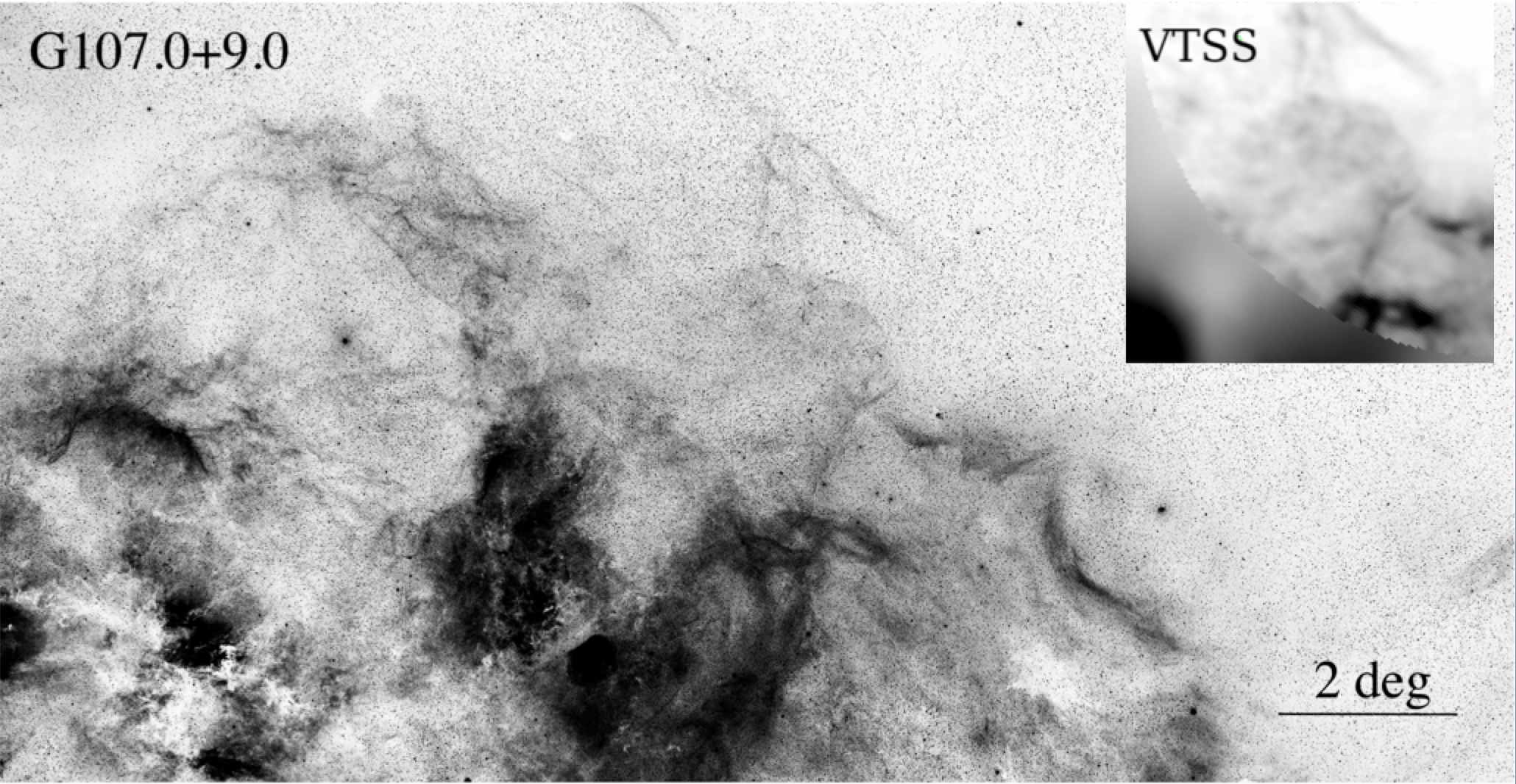} 
\caption{A $8\degr \times 16\degr$ wide-field composite MDW H$\alpha$ image centred on G107.0+9.0, with the VTSS H$\alpha$ image shown in the upper right-hand insert. North is up, East is to the left. \label{WIDE_MDW}}
\end{centering}
\end{figure*}

The line ratio of
I([S II])/I(H$\alpha$) $\geq 0.4$
has proven to be a successful observational tool for identifying optical nebulae as SNRs in both Galactic and extragalactic settings
\citep{Blair1981,Dopita1984,Fesen1985,Leonidaki2013,Long2017}. Using this line ratio criterium, several new Galactic SNRs have been optically discovered in the last two decades (e.g.,
\citealt{Stupar2007,Fesen2010}) along with dozens of additional SNR candidates \citep{Stupar2008,Stupar2011,Boumis2009,Ali2012,Sabin2013}.
Although the optical emissions of many known and suspected Galactic remnants are faint and
fragmentary \citep{Stupar2007,Stupar2011}, a few situated well off the
Galactic plane have been found to exhibit large and extensive optical emission
shells such as SNR G70.0-21.5 \citep{Boumis2002,Fesen2015}.

Here we report on the identification of another large and high Galactic
latitude SNR through its optical emission.  Its discovery was prompted
by \citet{Yuan2013} who in their search for faint planetary nebulae called
attention to a large, circular emission feature visible in the deep Milky Way
H$\alpha$ survey of Virginia Tech Spectral-Line Survey (VTSS;
\citealt{Dennison1998,Fink2003}).  They suggested that this nebula at Galactic
coordinates  $l = 107.1$, $b = +9.0$ might be an unrecognized SNR. 

\begin{figure*}
\begin{centering}
\includegraphics[angle=0,width=\textwidth]{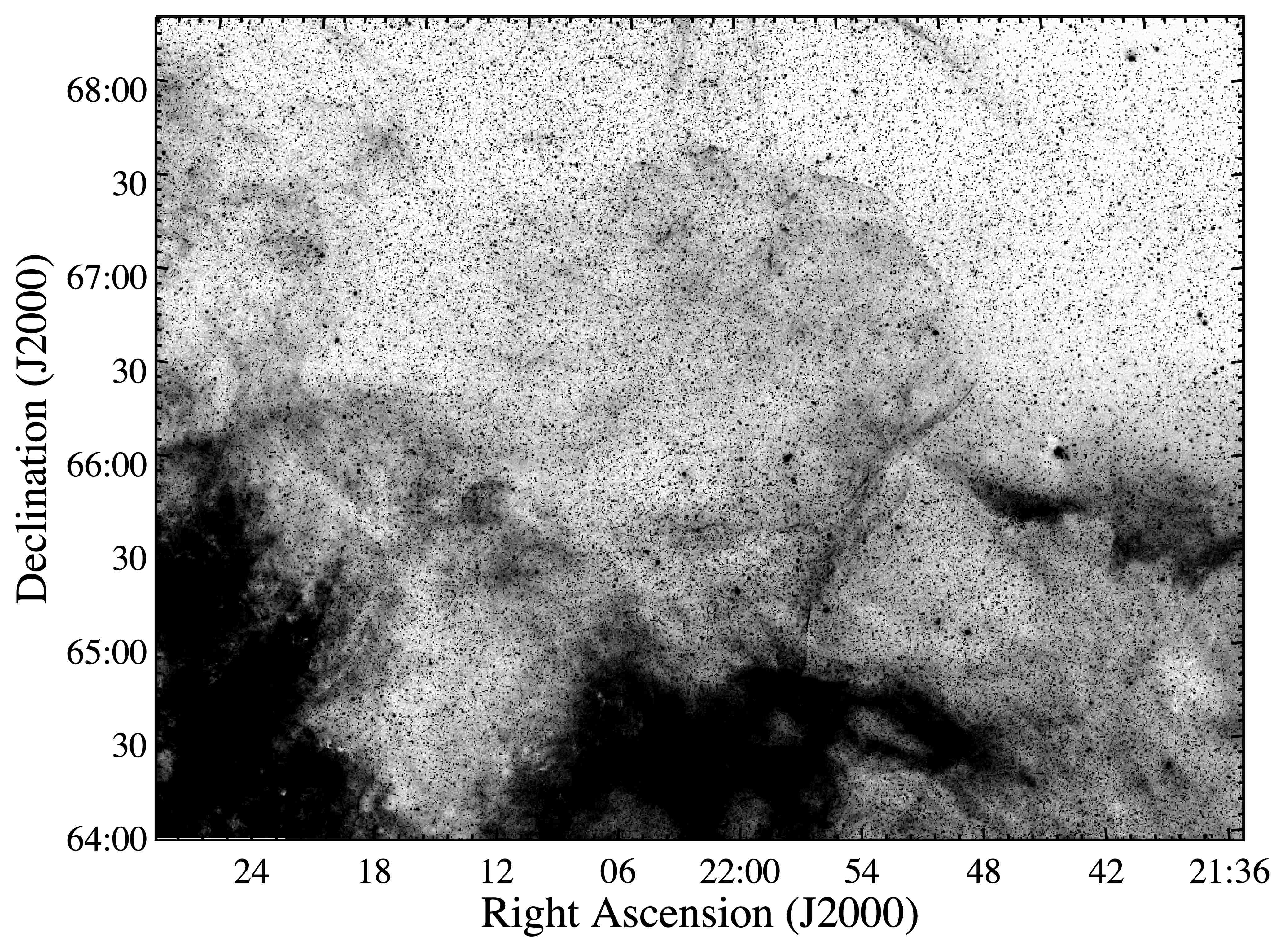}
\caption{MDW H$\alpha$ image of G107.0+9.0 
revealing numerous filaments along its northern, western and southwestern limbs. \label{MDW} }
\end{centering}
\end{figure*}

Many of the nebula's brightest filaments are visible on Digitized Palomar Sky
Survey (DSS1 and DSS2) images.  These include several filaments along it western and
southwestern limbs plus a thin filament along its northern limb. Even an
[O~III] 4959, 5007 \AA \  bright filament along the northwestern limb is easily visible in the blue DSS images. 

While there is no known SNR at this location in a recently
updated catalogue of Galactic SNRs \citep{Green2019} or in a listing of
high-energy known or suspected SNRs \citep{Safi2013}, the presence of so many
filamentary features suggested to us the likely presence of a large yet unknown
SNR situated well off the Galactic plane despite a lack of reported coincident
low-frequency radio emission.

In this paper, we present evidence that this nebula, G107.0+9.0, is a previously unrecognized Galactic SNR based on its optical emission properties and structure as determined from wide and narrow H$\alpha$,
[O~III] and [S~II] images plus low-dispersion, long-slit spectra.
Our imaging data and spectra are described in $\S$2, with results presented in
$\S$3.  The general properties of this new SNR including its shock velocities, distance, size and age are
discussed in $\S$4, with our conclusions summarized in $\S$5.

\begin{figure*}
\includegraphics[angle=0,width=16.5cm]{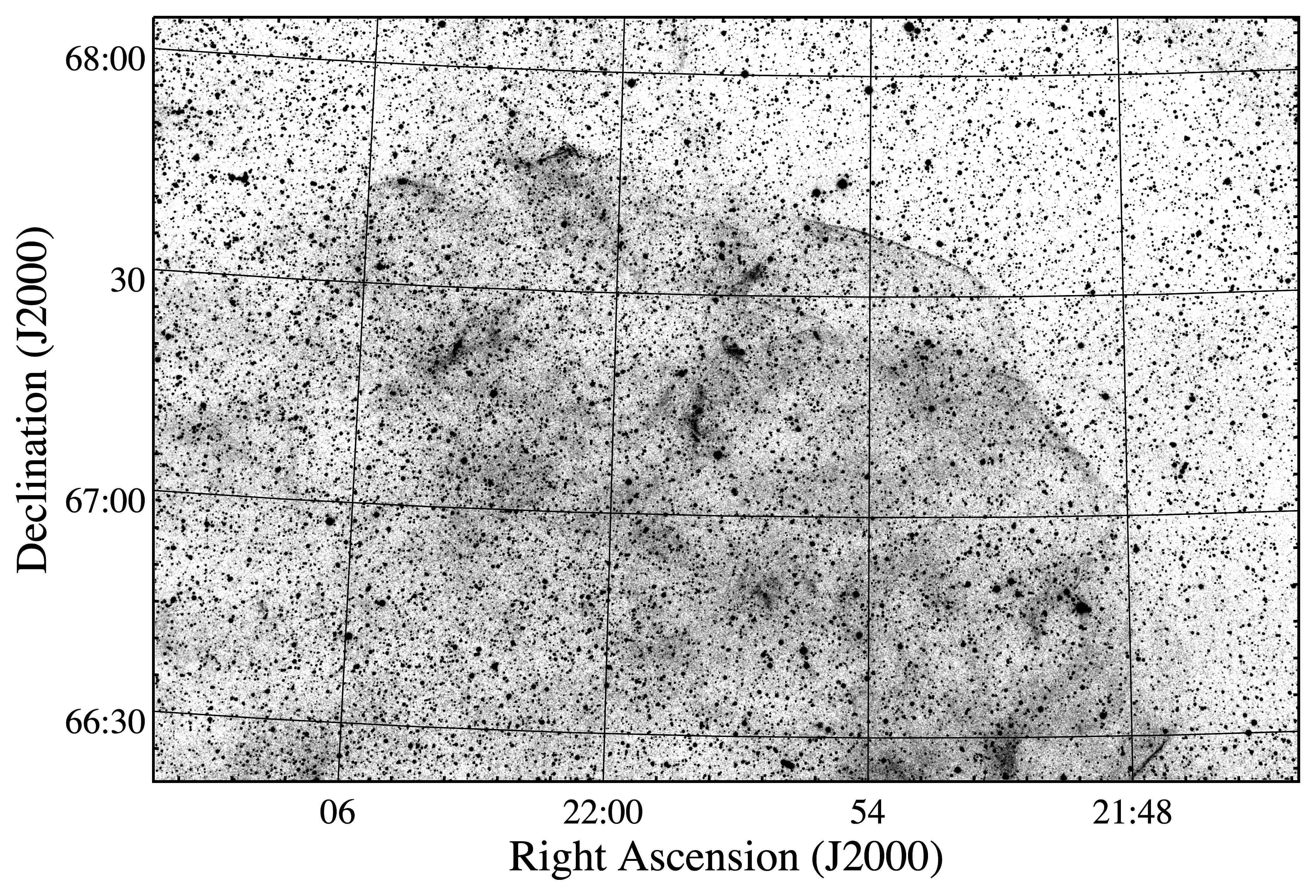} \\ 
\includegraphics[angle=0,width=16.5cm]{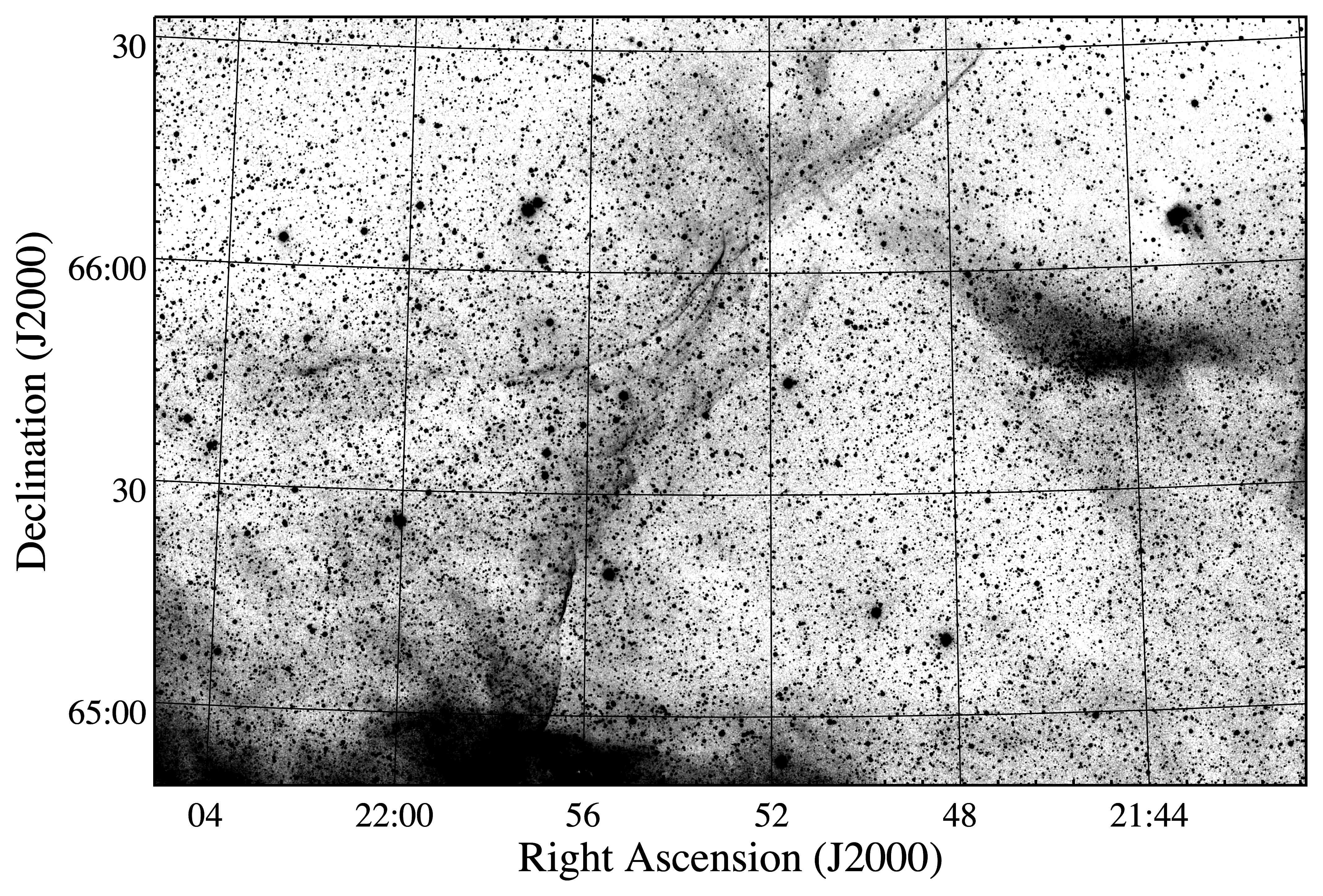}
\caption{Enlargements of MDW H$\alpha$ images of the northern and 
southwestern regions of G107.0+9.0. 
\label{MDW_NS}}
\end{figure*}

\begin{figure*}
\centering
\includegraphics[width=\textwidth]{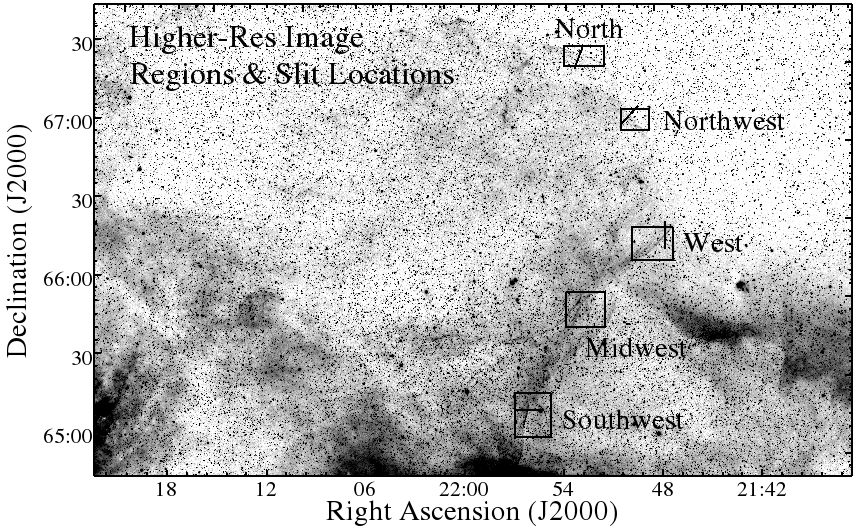}
\caption{MDW H$\alpha$ image of G107.0+9.0 showing locations
where higher resolution H$\alpha$ images were taken (five black boxes)
and the positions and orientations of four long slits (black lines) 
where spectra were obtained.  North is up, East is to the left. \label{Chart}}
\end{figure*}

\section{Observations}

Wide-field, low-resolution H$\alpha$ images of the region around the G107.0+9.0 nebula were obtained as part of the MDW Hydrogen-Alpha Survey\footnote{https://www.mdwskysurvey.org} in
November 2017. This survey uses a 130 mm telescope at the New Mexico Skies Observatory, with a FLI ProLine 16803 CCD and a 3 nm filter centered on H$\alpha$.  This telescope-camera system has a 
field-of-view (FOV) of approximately $3\fdg5\times3\fdg5$ with a pixel size of $3\farcs17$. Each field was observed with 12 $\times$ 1200 s exposures. 

Additional wide-field H$\alpha$, [O~III] and [S~II] images were obtained using two different setups. An Apogee U16M CCD attached to a 0.37 m telescope located in southern France at the Remote Observatories Southern Alps (ROSA) facility was used with 5 nm H$\alpha$ and [S~II] filters to provide images with a FOV $\simeq45'$.
Wide-field images were also obtained using a
Nikon F/2.0 200 mm focal length lens in front of an $2000 \times 3000$ pixel SBIG STL6303e CCD yielding a scale of $9.3''$ pixel$^{-1}$ and a FOV of $5.3 \times 8.0$ degrees. A series of $5 \times 2400$ s and $9 \times 1200$ s exposures were obtained through 3 nm [O~III] and 5 nm H$\alpha$ narrow passband filters, respectively.

Follow-up, higher-resolution images were obtained in September and October 2019 with the 2.4m Hiltner telescope at the MDM
Observatory at Kitt Peak, Arizona using the Ohio State Multi-Object
Spectrograph (OSMOS; \citealt{Martini2011}) in direct imaging mode. 
OSMOS uses a 4k $\times$ 4k CCD providing an effective FOV of
$18^{\prime} \times 18^{\prime}$. With $2\times2$ on-chip binning, this yielded
an image scale of $0.55^{\prime\prime}$ pixel$^{-1}$.

A series of 8 nm wide H$\alpha$ + [N~II] and [O~III] filter images using this set-up were taken with three exposures of either 900 s or 1200 s for each of five regions of G109.0+9.0. (Note: We will sometimes refer to these H$\alpha$ + [N~II] images as simply H$\alpha$ images since this nebula's [N~II] lines are typically weak.) These include two northern and northwestern regions plus three regions along the remnant's filament rich western limb.

Based on these images, low-dispersion optical spectra of five filaments in the
remnant's northern, western and southern regions were subsequently obtained
again with the MDM 2.4m Hiltner telescope using OSMOS now in spectroscopic
mode. Using a blue VPH grism (R = 1600) and a 1.2 arcsec wide slit, 
exposures of $2
\times 1200$ s and $2 \times 2000$ s were taken of covering 4000--6900 \AA \
with a spectral resolution of 1.2 \AA \ pixel$^{-1}$ and a FWHM = 3.5 \AA. Slit
orientations varied as a function of filament morphology and position angle.
Spectra were extracted from regions clean of appreciable emission along each of
the slits. Because two spectra taken of a bright SW
filament showed nearly identical line emissions, we present the spectra only of
one of these.

MDM images and spectra were reduced using the OSMOS reduction pipelines\footnote{https://github.com/jrthorstensen/thorosmos} in Astropy. Spectra were further reduced using PYRAF\footnote{PYRAF is a product of the Space Telescope Science Institute,
which is operated by AURA for NASA.} and L.A. Cosmic \citep{vanDokkum2001} to remove cosmic
rays and calibrated using a HgNe lamp and spectroscopic standard
stars \citep{Oke1974,Massey1990}.

\section{Results}

In order to place G107.0+9.0 (hereafter also
referred to as simply G107)  in context relative to other nebulosities along
and several degrees above the Galactic plane, Figure~\ref{WIDE_MDW} shows  an
approximately $8\degr \times 16\degr$ composite image made from some twenty MDW images. 
The G107.0+9.0 nebulosity appears as a nearly emission filled spherical nebula
situated almost three degrees above much brighter Galactic emission nebulae to its south and several degrees west of the large emission shell centred on the bright K0~III star, iota Cephei.  While the VTSS H$\alpha$ image insert in  Figure~\ref{WIDE_MDW} suggests G107's 
structure is nearly completely spherical toward the east, its 
extent to the south is uncertain due to coincident bright H~II region emissions to its south and southeast.
  
Greater detail can be seen in Figure~\ref{MDW} which shows a $4.4 \times
5.8$ degree composite MDW H$\alpha$ survey image.
While the VTSS survey image suggested a fairly complete circular emission
structure especially along G107's eastern limb, both data sets detect a rather
large nebula $\simeq$ 2.8 degrees in diameter roughly centered at RA (J2000) =
$22^{\rm h}00^{\rm m}25^{\rm s}$, Dec (J2000) = 66${\degr}26'$ corresponding to
Galactic coordinates of $l = 107.0, b = 9.0$ degrees. 

As shown in the enlargements of the MDW H$\alpha$ images presented in
Figure~\ref{MDW_NS}, G107 possesses numerous filaments along its northern
and western boundaries, plus considerable interior diffuse emission 
especially in its western half and 
along its northern and western limbs.  While only one bright filament is
readily noticeable along G107's northernmost limb (upper panel), many such
filaments can be seen along its western and southwestern edge (lower panel).

These wide FOV and relatively low-resolution MDW images, deep as they
are, hide a far richer, more complex and quite extensive network of optical filaments visible in
higher resolution images. Below we will present and briefly discuss our higher
resolution images for five regions around G107 in clockwise order starting
with the thin, bright northern filament (see Fig.~\ref{Chart}).    

\begin{figure*}
\centering
\includegraphics[width=\textwidth]{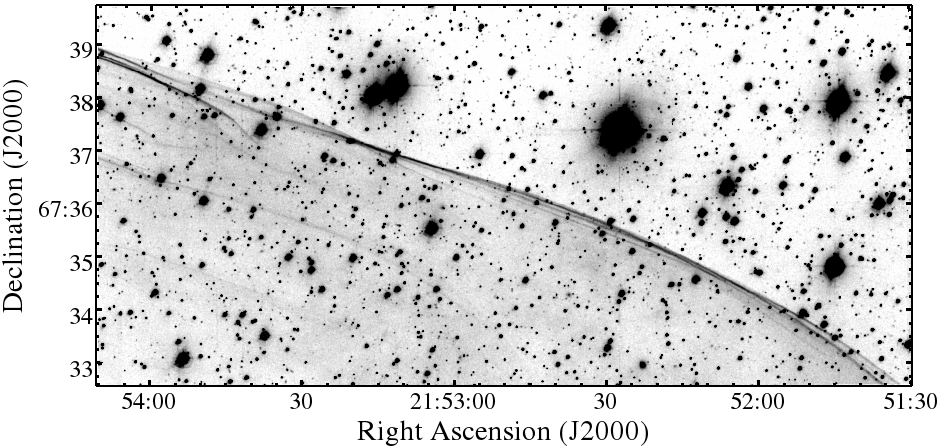}
\caption{H$\alpha$ image of a sharp Balmer 
dominated filament along the north-central limb of G107.0+9.0. 
\label{Balmer}}
\end{figure*}

\begin{figure*}
\centering
\includegraphics[width=\textwidth]{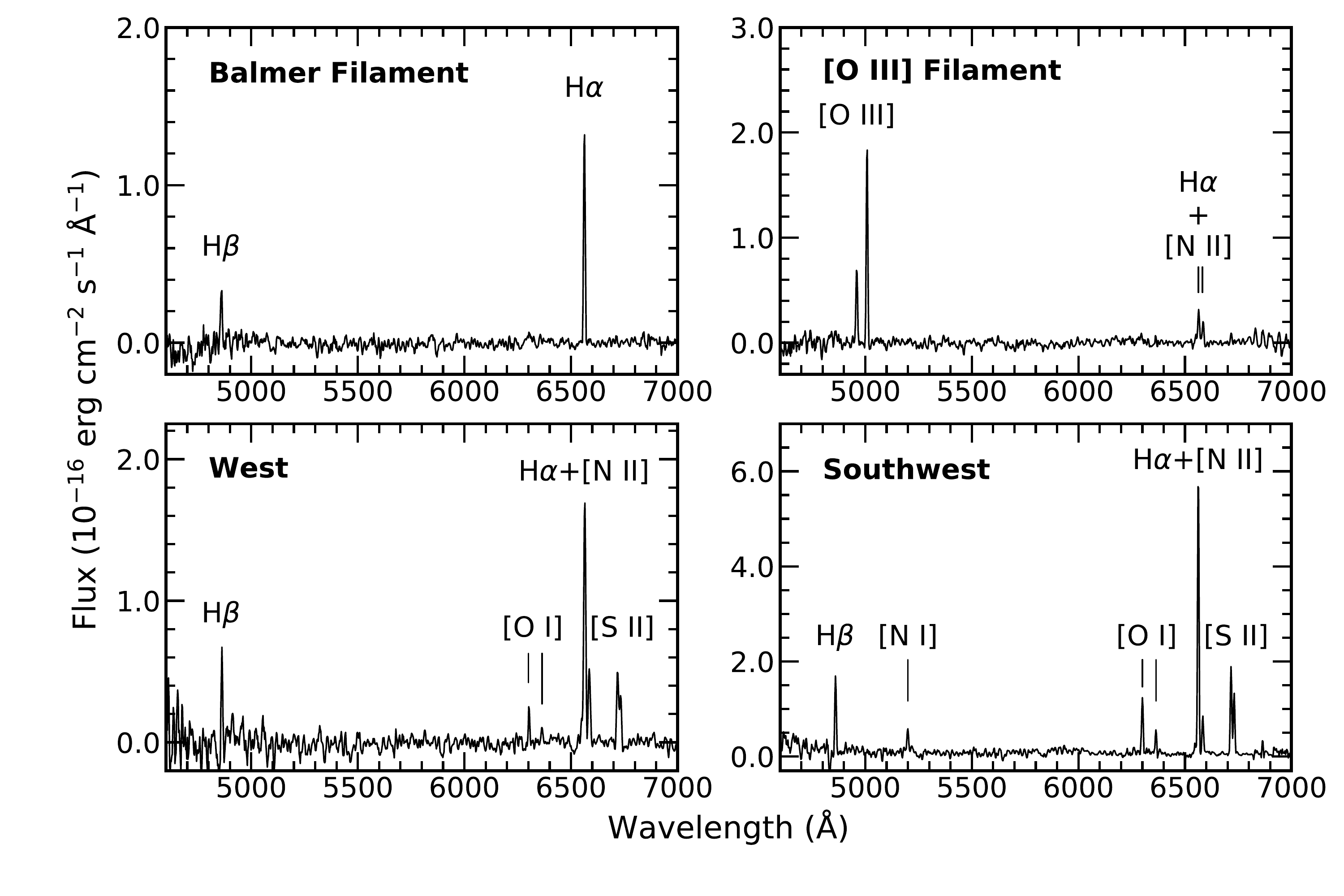}
\caption{Optical spectra of four regions in the G107.0+9.0 SNR. 
See Figure~\ref{Chart} for slit locations. \label{spectra}}
\end{figure*}

{\bf{North Filament:}} A sharp and bright filament seen prominently in Figure~\ref{MDW_NS}
(upper panel) along G107's northern limb, and even easily visible on DSS images, is shown
better resolved in Figure~\ref{Balmer}.  Appearing as a textbook
example of an optically emitting interstellar shock front, this filament consists of
several closely overlapping shocks view edge-on lying in the plane of the sky.
Considerable interior trailing emission is readily visible, making for a distinct emission
boundary of shocked and unshocked interstellar medium. 

Spectra of this filament, shown in the  upper left panel in Figure~\ref{spectra},
revealed a Balmer dominated, non-radiative type of shock emission. No emission
lines other than H$\alpha$ and H$\beta$ were detected in the wavelength region
$4600 - 7000$ \AA. This included absence of the [S~II] 6716, 6731 \AA \
lines, commonly seen in the shock filaments of SNRs. 
An observed H$\alpha$/H$\beta$ ratio $\simeq$4.1 suggests an $E(B - V)$ $= 0.25
\pm 0.05$ assuming an intrinsic unreddened value of 2.76 for T $= 2\times 10^4$
K \citep{Osterbrock2006}.

Although such Balmer dominated spectra have most often been associated with
relatively high-velocity shocks ($\geq$ 1000 km s$^{-1}$) present in young
SNRs, they have also been seen in much older remnants where slower shock velocities
$\simeq 200 - 350$ km s$^{-1}$ are present; e.g., the Cygnus Loop
\citep{Raymond1983,Fesen1992,Medina2014} and G156.2+5.7 \citep{Gerardy2007}.

\begin{figure*}
\centering
\includegraphics[angle=0,width=0.9\textwidth]{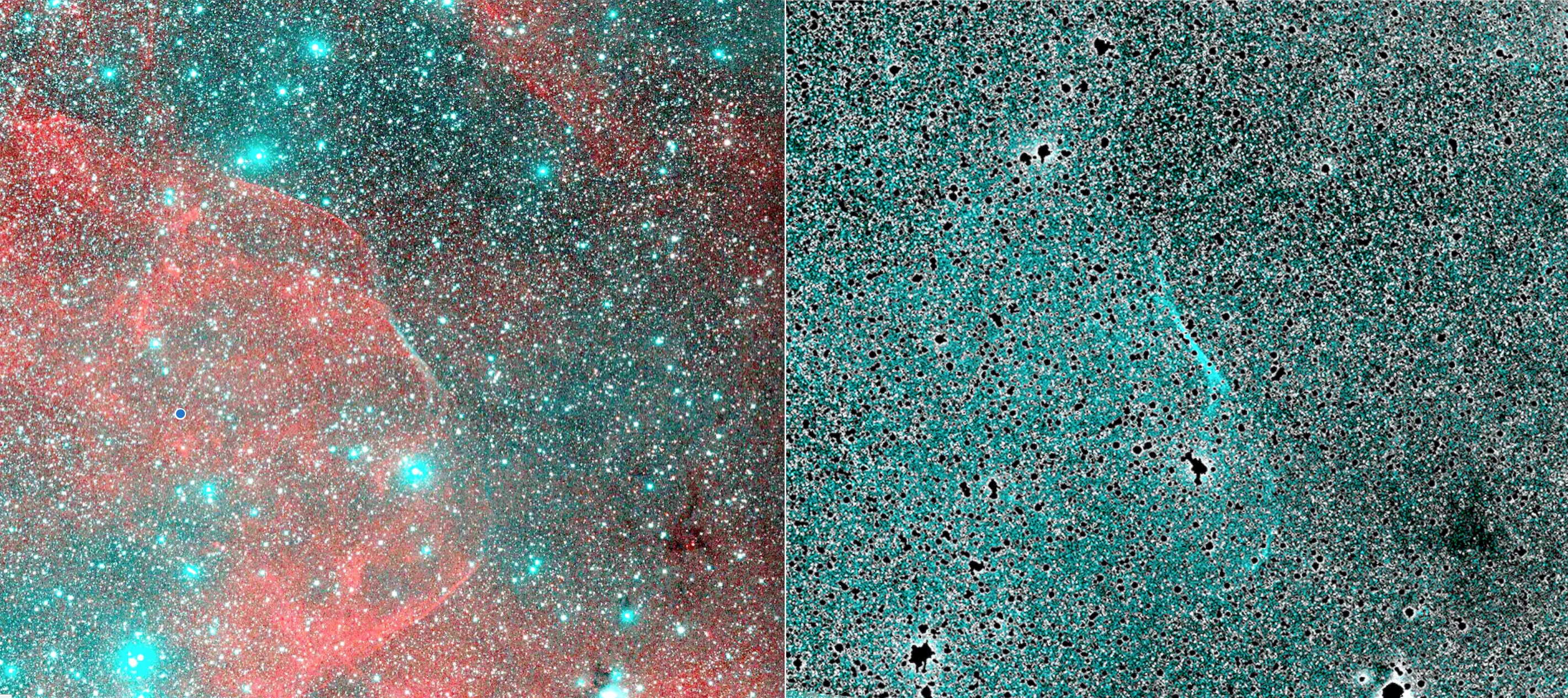}
\caption{Left: A color H$\alpha$ (red) and  [O~III] (blue) composite image of the remnant's northwestern region. 
Right: Continuum subtracted [O~III] image
of the same region showing the presence of just one bright [O~III] filament. North is up, East is to the left. \label{Hu_color} }
\end{figure*}

{\bf{Northwest Region:}} Matching wide-field H$\alpha$ and [O~III] images of
G107's northern and northwestern half revealed only a single [O~III] bright
filament along with faint diffuse interior [O~III] emission. These images are
shown in Figure~\ref{Hu_color} where the left panel shows a color composite
image where H$\alpha$ emission is red and [O~III] emission is blue. The
right hand panel of this figure shows a continuum subtracted [O~III]
emission image of this same region. As can be seen in the right panel, the
[O~III] emission filament is relatively broad and partially resolved with much
fainter filamentary and curved emission extending farther to the south and
west. 

\begin{figure*}
\includegraphics[angle=0,width=9.0cm]{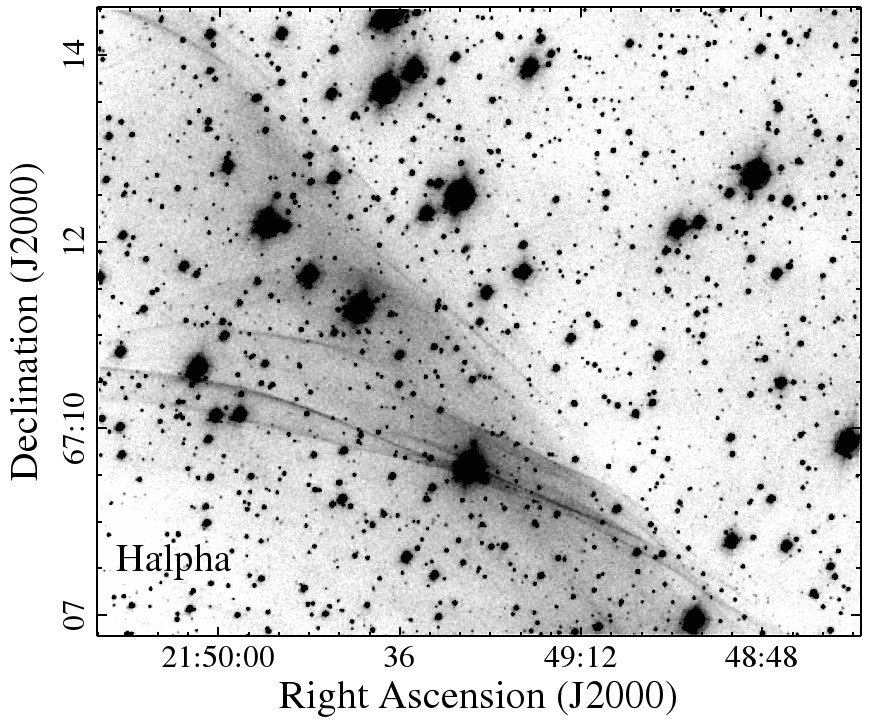}
\includegraphics[angle=0,width=8.1cm]{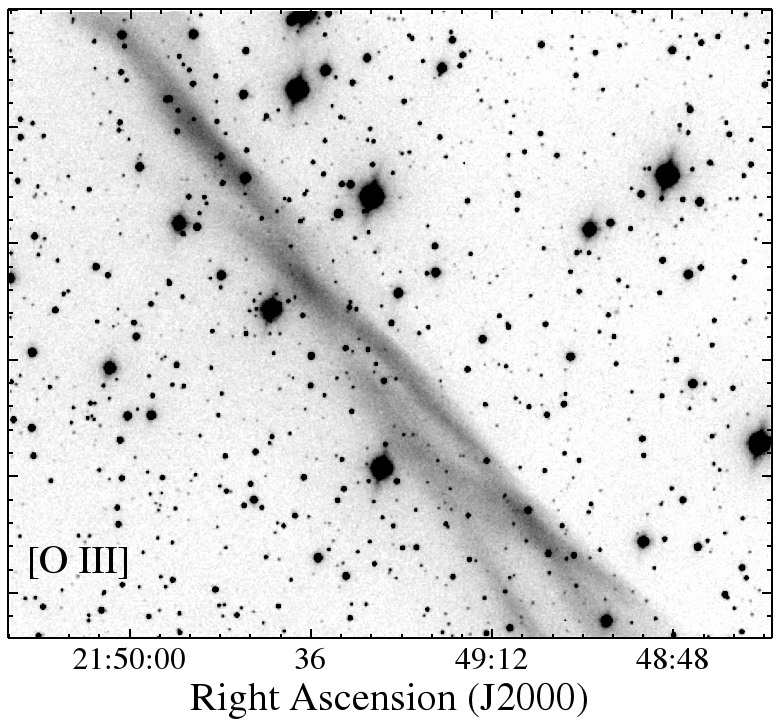}
\caption{H$\alpha$ and [O~III] images of the region
containing the remnant's only bright [O~III] emission filament. \label{O3_filament}  }
\end{figure*}

To explore the positional relation of H$\alpha$ and [O~III] emission along
this section of G107's NW limb, we obtained higher resolution H$\alpha$ 
and [O~III] images of this area (see Fig.~\ref{Chart}) which
are shown in  Figure~\ref{O3_filament}.  These images present a striking
morphological difference in the emission structure between H$\alpha$ and
[O~III].  The [O~III] emission is limited mainly to G107's outer limb
and is considerably more diffuse and limited in radial extent as compared to that seen in H$\alpha$.

\begin{figure*}
\centering
\includegraphics[width=0.9\textwidth]{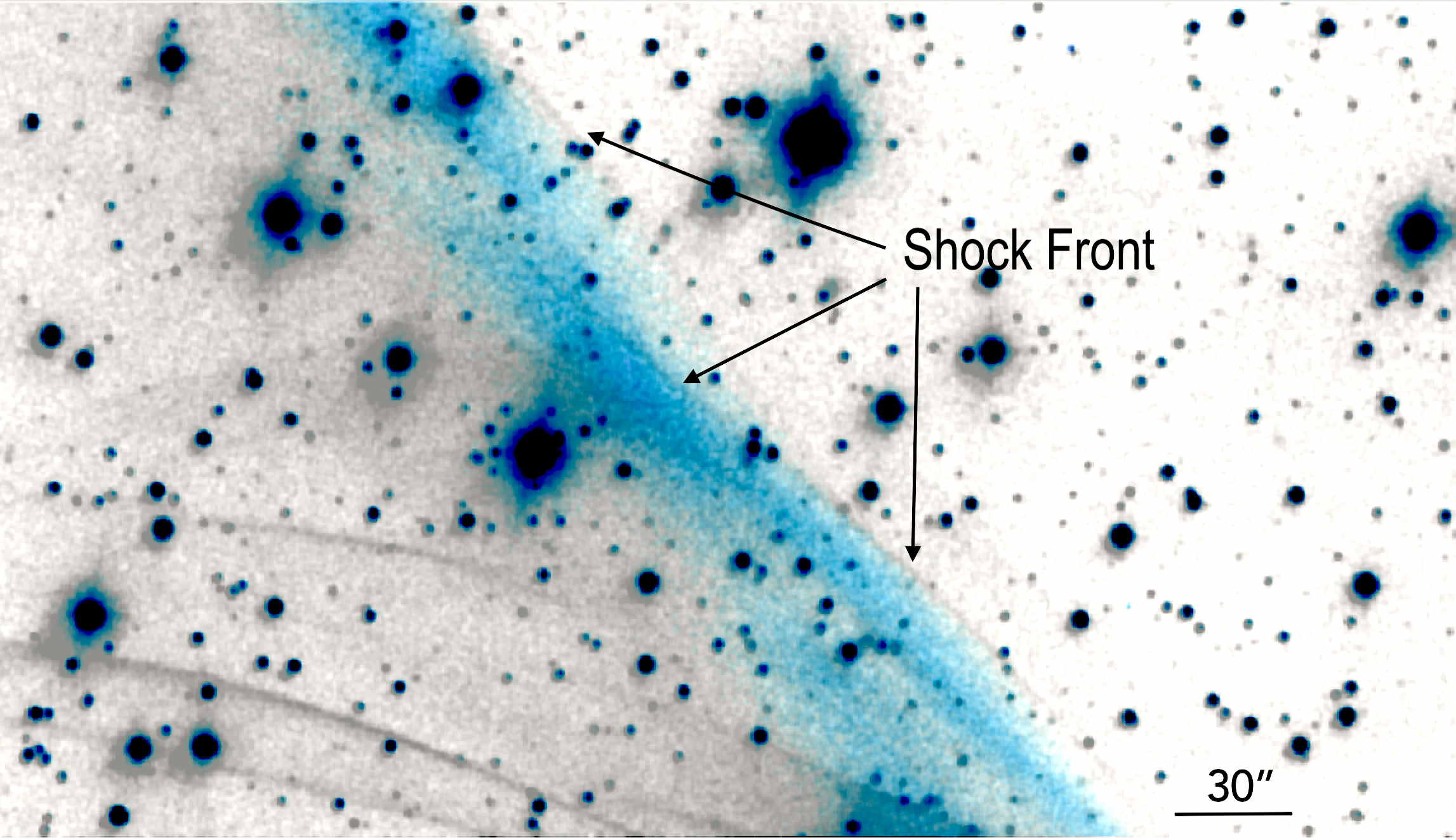}
\caption{Overlay of the [O~III] image (blue) on the 
H$\alpha$ image (grey) of the [O~III] filament region of
G107.0+9.0 showing the formation of increasing and then 
decreasing [O~III] emission with increasing distance behind 
faint H$\alpha$ filaments which mark the remnant's shock front.  \label{O3_color}}
\end{figure*}

\begin{figure*}
\centering
\includegraphics[angle=0,width=18.0cm]{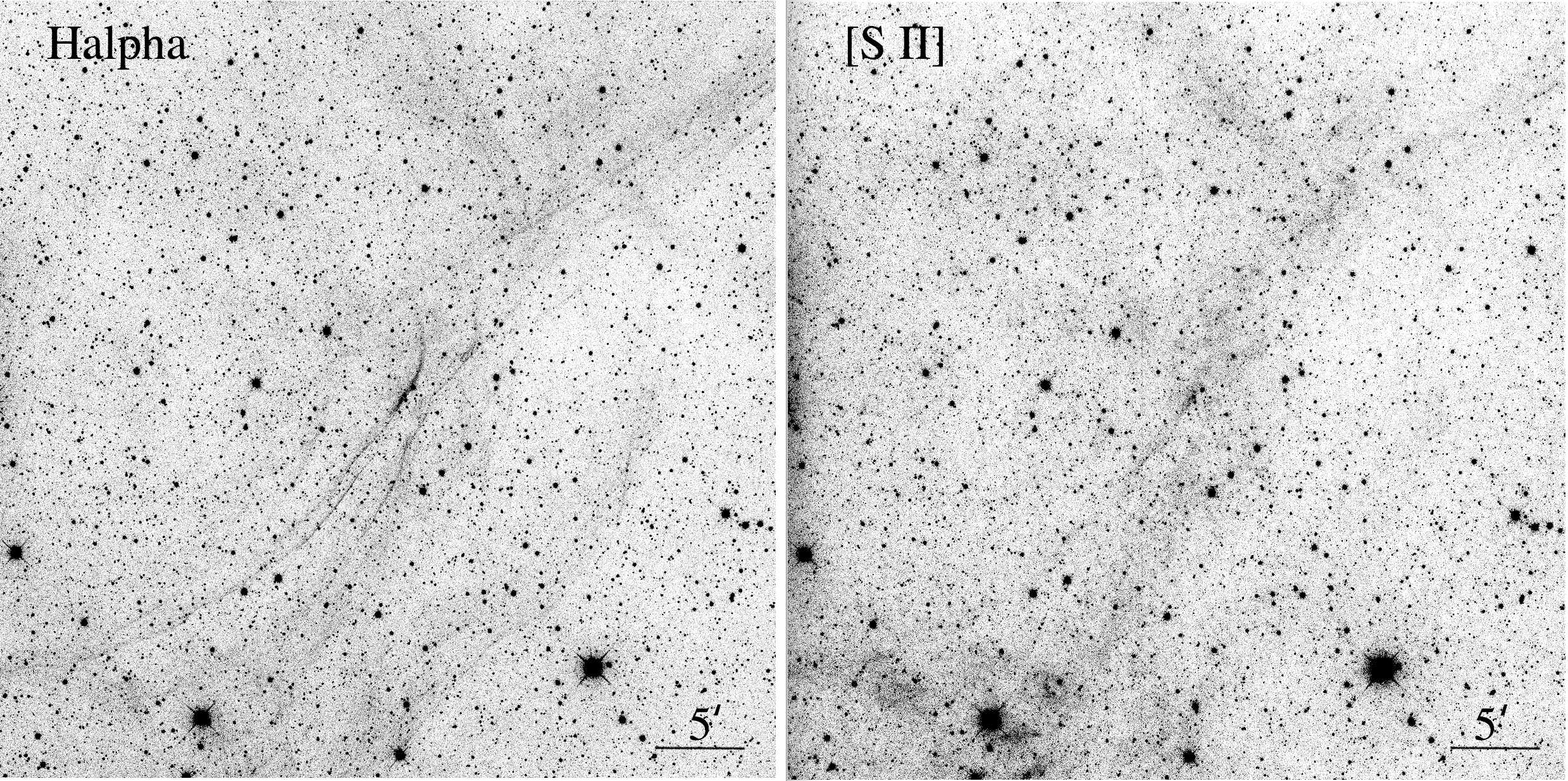}
\caption{Comparison of the remnant's 
H$\alpha$ emission (left) with that of its [S~II] emission (right)
along a portion of its western limb. 
Note the more diffuse appearance of the [S~II] emission and
the absence of corresponding sharp H$\alpha$ filaments. This
suggests the wide-spread presence of non-radiative Balmer filaments.
Images are approximately $44'$ square with center coordinates 21:53:07, +66:01 (J2000).
North is up, East is to the left. \label{Martins} }
\end{figure*}

\begin{figure*}
\centering
\includegraphics[height=0.32\textheight]{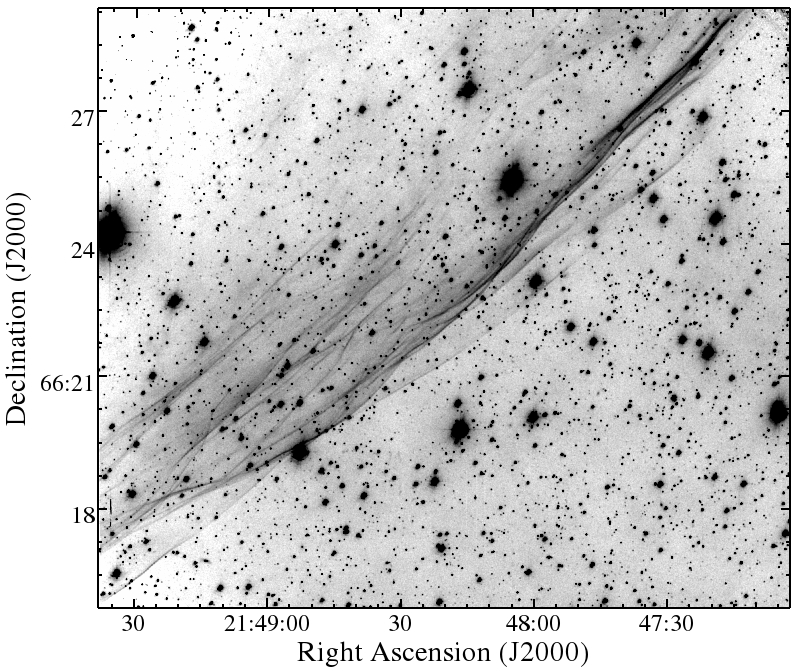}
\includegraphics[height=0.32\textheight]{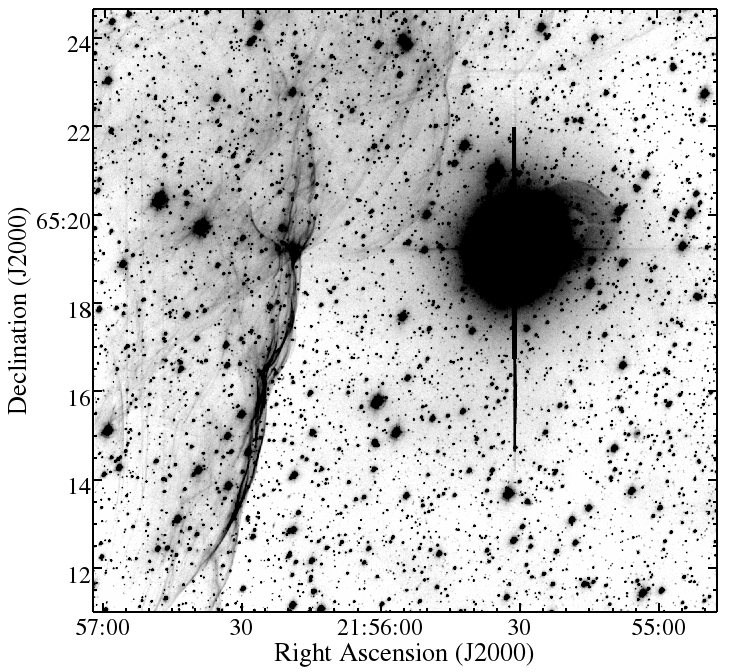} \\
\includegraphics[align=t,height=0.33\textheight]{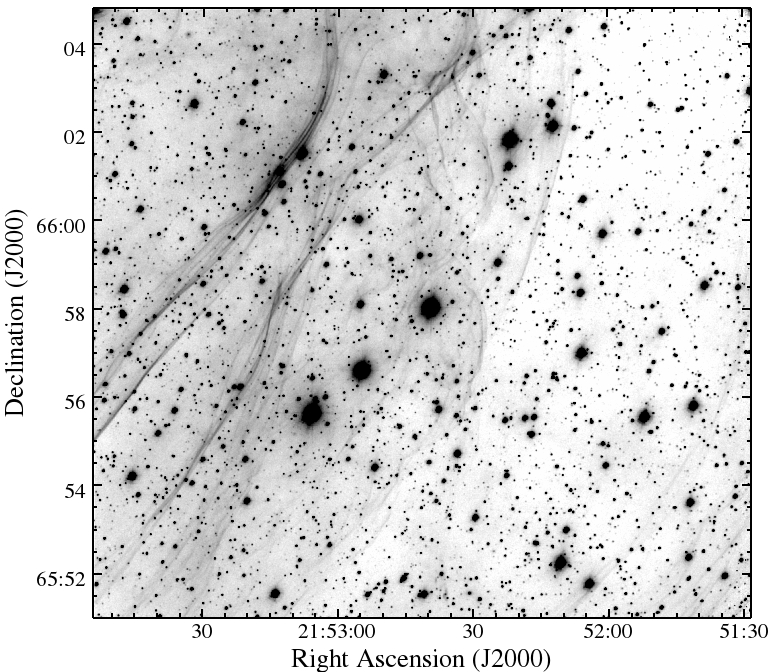}
\includegraphics[align=t,height=0.33\textheight]{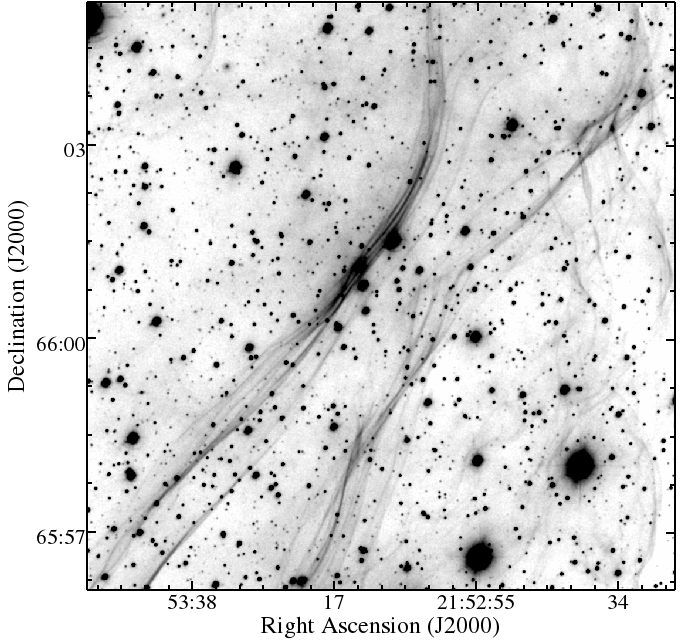}
\caption{Upper Panels: H$\alpha$ image of G107.0+9.0's Northwest (left) and Southwest (right) regions.
Lower Panels: H$\alpha$ image of G107.0+9.0's Midwest region (left) and a close-up of its filament rich area (right) displaying several near parallel filaments indicating the presence of multiple shock fronts. \label{West}}
\end{figure*}

A composite overlay of these H$\alpha$ and [O~III] images is shown in
Figure~\ref{O3_color} where H$\alpha$ emission is shown as grey with
[O~III] emission colored blue. Faint thin H$\alpha$ emission
filaments can be seen to lie immediately ahead of the diffuse [O~III]
emission.  The thin H$\alpha$ filaments would appear to mark the location of
G107's expanding shock front followed by a broad recombination zone where
[O~III] emission gradually strengthens and then gradually fades with
increasing distance away from the shock front. The thickness of this apparent
[O~III] recombination zone perpendicular to the faint H$\alpha$ filaments
is $\sim 20'' -  25''$. Because we only imaged this one section of the [O~III]
bright filament, we cannot say anything about this H$\alpha$ vs [O~III]
emission structure farther to the southwest where the [O~III]
filament appears to split.

A long slit spectrum cutting across the [O~III] filament with the $1.2''
\times 10'$  slit placed perpendicular to the  [O~III] filament's long
axis was obtained. The spectrum for the thin faint H$\alpha$ filament out ahead and bordering the [O~III] emission showed only H$\alpha$ and H$\beta$ indicating it is a non-radiative emission filament. 

The spectrum for the [O~III] bright filament region is shown in the upper right-hand
panel of Figure~\ref{spectra}.  As expected, [O~III] emission dominates
the spectrum, with an [O~III] 5007/H$\alpha$ ratio of $5.5 \pm 0.3$. This
value is near the upper range of values seen generally in old, well-evolved SNRs
\citep{Fesen1985,Fesen1996}, and is common in post-shock cooling emission
filaments close to the shock front as marked by non-radiative emission
filaments \citep{Hester1994}.  This is consistent with that shown in the
displacement of the [O~III] behind the faint H$\alpha$ filaments shown in
Figure~\ref{O3_color}.  An analysis of this H$\alpha$ and [O~III] spatial displacement is discussed in $\S$4.

{\bf{West Limb Regions:}} It was clear from even DSS and MDW images
that G107 exhibits many long, curved filaments along the whole of its western
and southwestern limbs.  Interestingly, however, this is not the case when viewed in
[S~II] 6716, 6731 \AA \ line emission. 

As shown in the wide-field images
presented in  Figure~\ref{Martins}, G107's [S~II] emission is far less
filamentary and much more diffuse in appearance than that seen in H$\alpha$. 
Although  [S~II] emission is present in the areas where there is strong
H$\alpha$ emission, there is poor one-to-one correspondence in many places, 
and several long thin  H$\alpha$ filaments have no obvious [S~II] counterparts.

This suggests these [S~II] absent emission filaments are non-radiative, Balmer dominated shock
filaments much like the one seen along G107's northern limb discussed
above.  Some of the clearest examples are the two thin filaments extending from
the center bright filamentary H$\alpha$ emission feature toward the southeast
(lower left) and off the image's left-hand edge. 

Figure~\ref{West}  presents our higher resolution
H$\alpha$  images of three sections of G107's western limbs; namely, 
West, Midwest, and Southwest regions (see Fig.~\ref{Chart}). As
can be seen in these images, G107 possesses a fairly complex 
arrangement of nearly aligned filaments
along its western limb. The amount of detail in these images indicating multiple shock fronts is impressive 
and we have tried to show some of this complex structure in 
the lower right-hand panel of Figure~\ref{West} where we present an
enlargement of one small section along G107's midwestern limb. Closely spaced and nearly parallel filaments
can be seen to merge or diverge in a complex 
arrangement.  It is interesting to note that these brighter western regions mark only part of the
remnant's full western extent since a series of much fainter, farther western filaments
can be seen in Figures~\ref{MDW_NS} and \ref{Martins}.

Spectra of two filaments along G107's western
limb (see Fig.~\ref{Chart}) are shown in Figure~\ref{spectra}.
Both filament regions sampled show similar emission lines, with  no [O~III] emission seen and
[S~II]/H$\alpha$ ratios of $0.51 \pm 0.03$ for the West filament, and $0.53 \pm
0.02$ for the Southwest filament. These line ratio values are well within the range seen
in radiative shock filaments of SNRs \citep{DD1983,Dopita1984,Fesen1985}.  This, together
with the presence of strong [O~I] 6300, 6364 \AA \ line emissions,
confirm these filaments to be shock-heated SNR type emissions. The density sensitive 
[S~II] 6716,6731 line ratio for both filaments is near the low density limit of $\simeq1.4$ indicating a post-shock filament density $\leq100$ cm$^{-3}$.

The observed H$\alpha$/H$\beta$ ratio for the Northwest filament was 3.8, while
spectra of West and Southwestern filaments ranged from $4.0 \pm 0.15$ to $4.6 \pm 0.25$.
These values suggest a $E(B-V)$ range of $0.23 - 0.32$, similar to that found for the
northern Balmer dominated filament. 

\section{Discussion}

The imaging and spectral data presented above leave little doubt that the
G107.0+9.0 nebulosity is a previously unrecognized Galactic supernova remnant. This conclusion is supported by its near spherical shape, the presence of both radiative and non-radiative filaments found throughout the nebula, and the
coincidence of sharp H$\alpha$ filaments immediately ahead of
[O~III] bright emission along its NW limb. 
Several of the remnant's brighter filaments show
characteristic SNR lines ratios including [S~II]/H$\alpha$ ratio $>0.4$,
widely used as a diagnostic for shocked gas.  Moreover, these filaments also show strong
[O~I] 6300, 6364 \AA \ line emission which is a secondary
indicator for shock emission and is commonly observed in evolved SNRs
\citep{Fesen1985,Kop2020}. 

Morphologically, G107's filament rich optical emission structure is also
consistent with a SNR nature. Its numerous and delicate filaments
seen along its northern and western boundary are matched by few other Galactic remnants. Dozens of long and gently curved filaments indicate the remnant's shocks are expanding into large-scale, local interstellar regions with varying preshock densities.

Given G107's extensive and relatively bright optical structure, it is a bit
surprising that it was not identified as a SNR until now given the number of previous
optical Galactic SNR searches. This is especially surprising given its visibility on the Digital Sky Survey images.  While few of its filaments are
particularly bright, the sheer number of its filaments and extended diffuse emissions give G107 a relatively high total optical luminosity.

\begin{figure}
\centering
\includegraphics[width=\columnwidth]{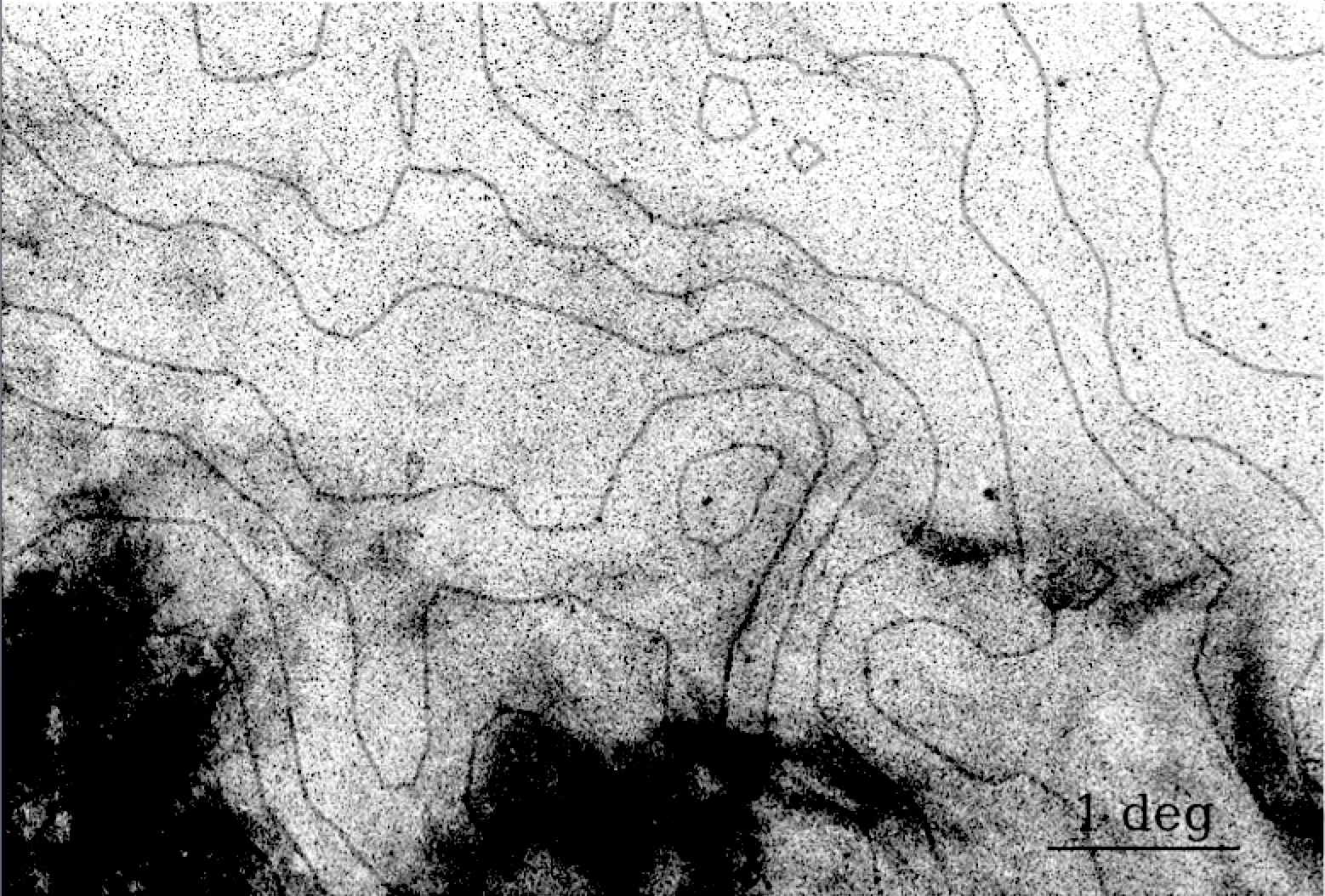}
\caption{Bonn 1420 MHz radio emission contours overlaid on to the MDW H$\alpha$ image of G107.0+9.0.
North is up, East is to the left. \label{radio}}
\end{figure}

With a diameter of nearly three degrees, G107 ranks among the largest angular size remnants  
of the 297 currently confirmed Galactic SNRs \citep{Green2019}. Centred at a Galactic latitude of nine
degrees, it is also among the farthest from the Galactic plane. Given this location, it is perhaps not too surprising that G107 has been missed in radio SNR surveys which have primarily limited their focus within
$\simeq 5$ degrees of the Galactic plane
\citep{Gerbrandt2014,Anderson2017,Gao2011,Gao2014,Gao2020}, although there have
been a few radio discoveries at much high latitudes \citep{Kothes2017}. 

No clear associated infrared emission features with G107 are seen in either WISE or IRAS infrared images and no associated X-ray emission is visible in ROSAT images. However, some relatively faint 1420 MHz radio emission \citep{Reich1982} appears coincident along its western limb,  with
emission contours following the remnant's optical emission. This is shown in Figure~\ref{radio}. 

The spectrum of the [O~III] bright filament along G107's NW limb showing an [O~III] 5007/H$\alpha$ ratio $\simeq5$ indicates a shock velocity of at least $100$ km s$^{-1}$ \citep{Raymond1979,Shull1979,Allen2008}. In contrast, the lack of any appreciable [O~III] emission in the filaments along G107's western limb (Fig.~\ref{Hu_color}) or seen in the spectra of the two western filaments (Fig.~\ref{spectra}) indicates shock velocities at or below 70 km s$^{-1}$.  A slower shock speed along G107's western edge compared to its northern limb might help explain the eastern indentation seen in the remnant's brightest western emission.

\begin{figure}
\centering
\includegraphics[width=0.45\textwidth]{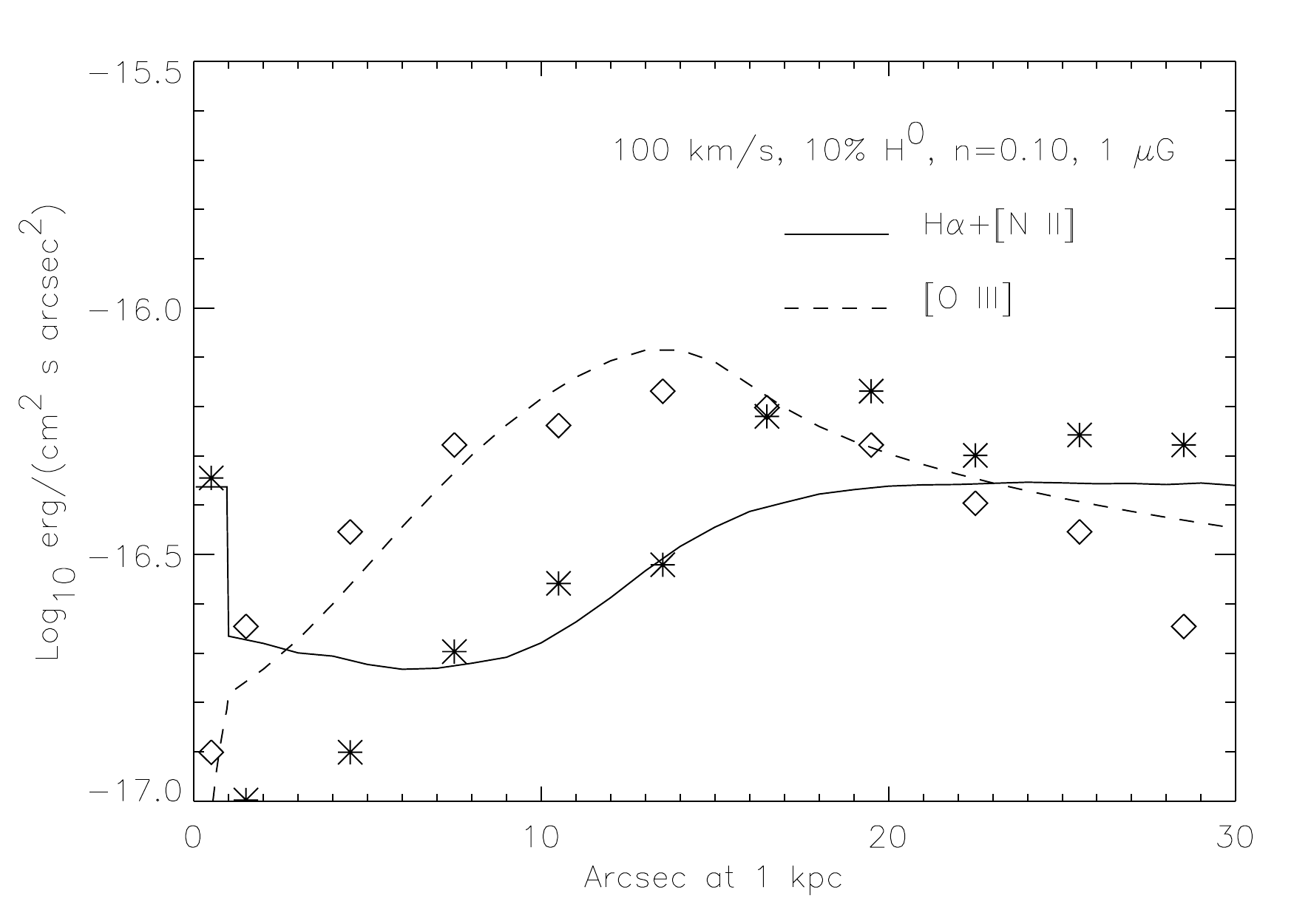}\\
\includegraphics[width=0.45\textwidth]{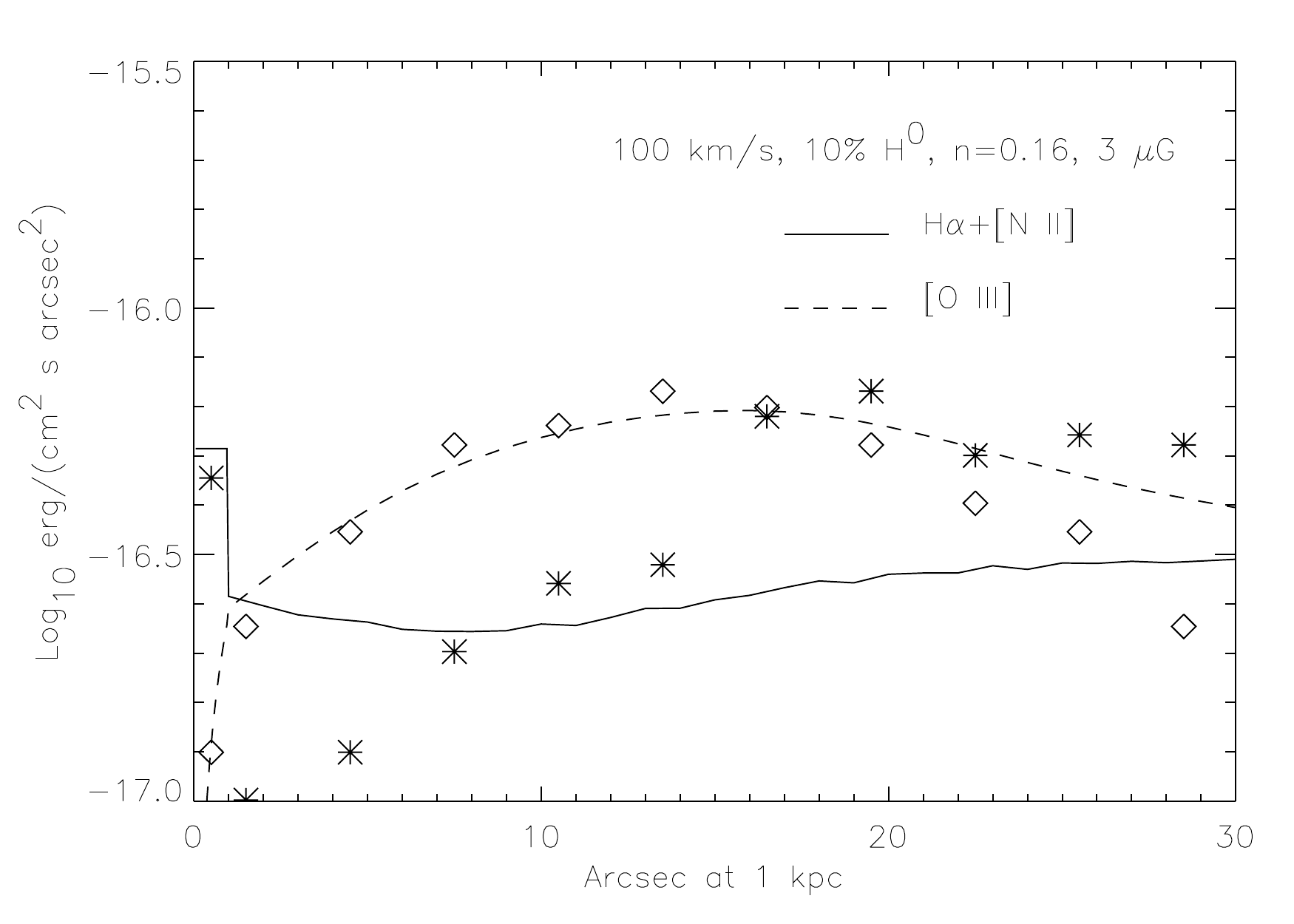}
\caption{Model predictions of H$\alpha$ + [N~II] and [O~III] emissions 
as a function of distance behind a 100 km s$^{-1}$ shock. The asterisks and diamonds are measurements from our 
H$\alpha$ + [N~II] and [O~III] images, respectively.
\label{Raymond}}
\end{figure}

While the distance to the G107 remnant is unknown, we can set some limits on it.
Its large angular size
(dia. $\simeq 50$ pc kpc$^{-1}$) and shock velocities below 70 km s$^{-1}$ for much of its western limb are suggestive of an old, physically large and well evolved SNR likely in its post-Sedov, radiative phase of evolution.
However, at its Galactic latitude of nine degrees, it is unlikely to be much farther than about 2 kpc, because distances above that it would have a physical diameter greater 100 pc thereby rivaling the largest known Galactic SNRs plus a location in excess of 300 pc off the plane.

Some insight into G107's likely physical size and hence to its distance can be gained from a comparison of other large optical remnants located far off the Galactic plane like G107. The Cygnus Loop remnant also lies well off the Galactic plane at $b$ = $8.5\degr$, is similar in angular size to G107 (ignoring the Cygnus Loop's southern blowout region), is also expanding into a fairly low density region ($n_{\rm 0}$ = 0.4 cm$^{-3}$), and has an estimated diameter and distance of 40 pc at a distance of 0.74 kpc, respectively \citep{Fesen2018}. However, the Cygnus Loop is likely
much younger than G107 (age $\sim 20 \times 10^{3}$ yr) and is still firmly in its Sedov phase of evolution.

\begin{figure*}
\centering
\includegraphics[width=0.48\textwidth]{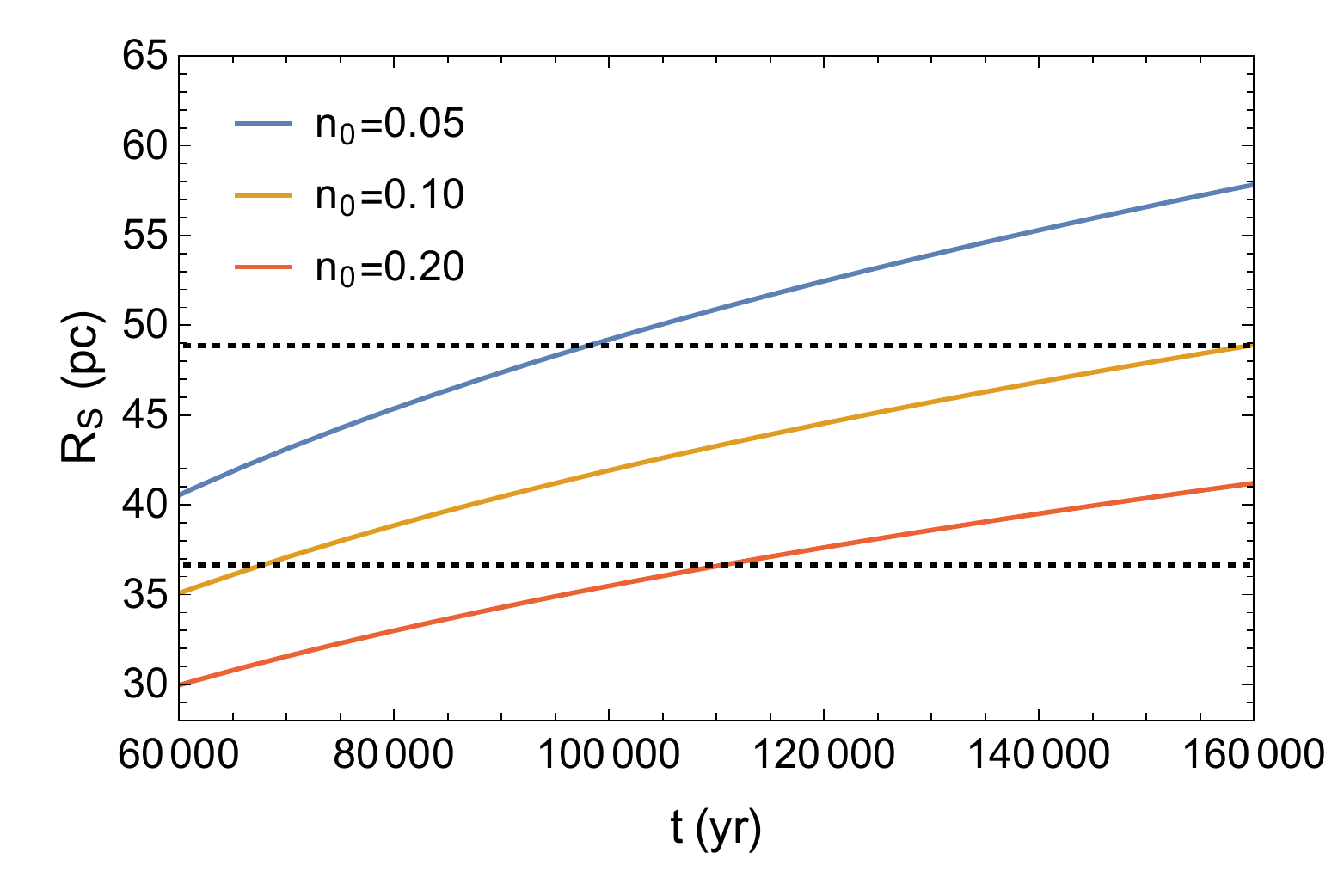}
\includegraphics[width=0.48\textwidth]{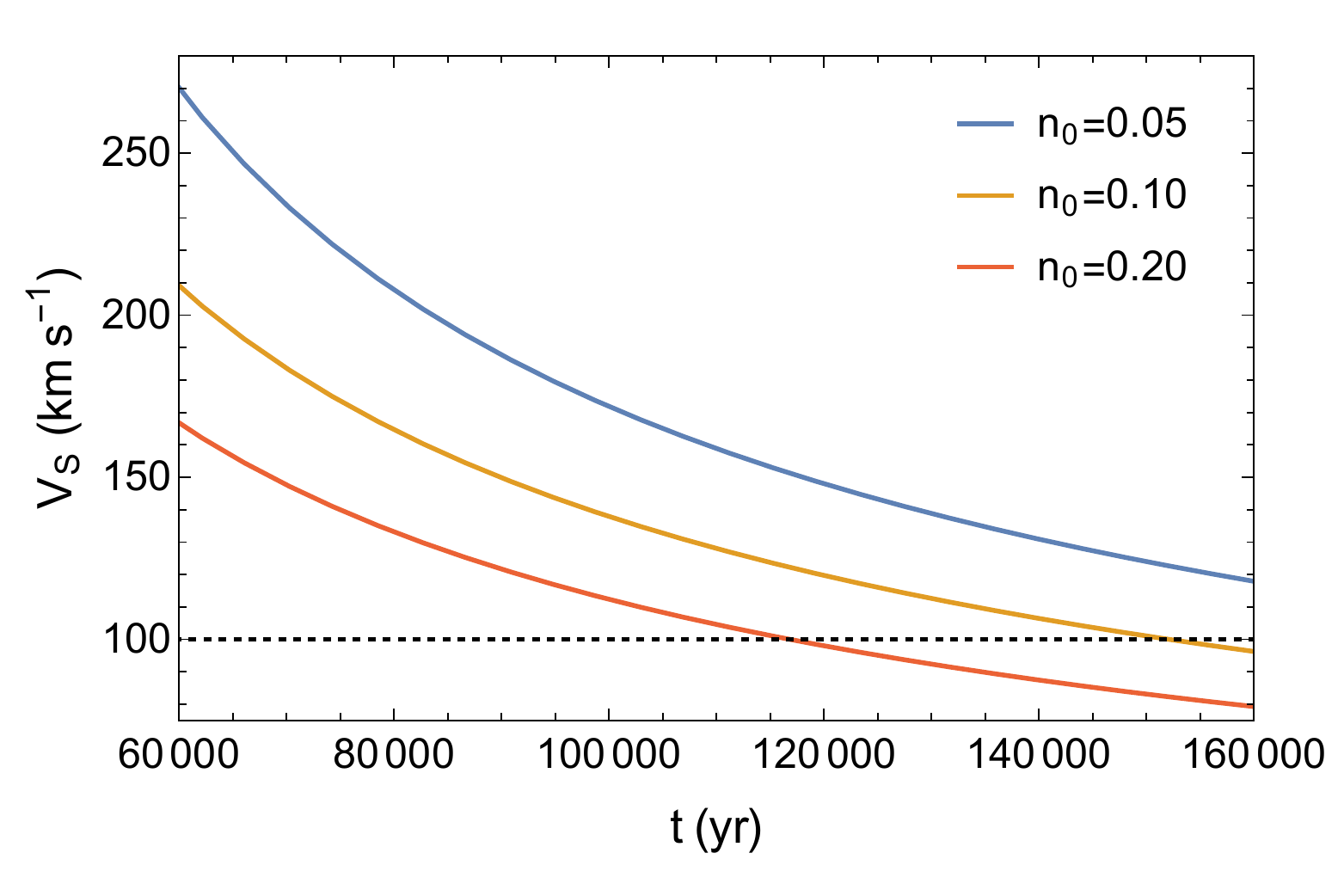}
\caption{Radius and velocity as a function of time for three ambient densities from \citet{Cioffi1988}.  An explosion energy of $0.5\times10^{51}$ erg was assumed. On the left the minimum and maximum radius are marked by the dotted lines by using a remnant distance of 1.5 kpc (R = 37 pc) or 2 kpc (R = 49 pc), respectively. On the right the observed shock velocity of the [O~III] filament is marked with a dotted line.
\label{CioffiModels}}
\end{figure*}

Perhaps a  better comparison is the Monoceros Loop SNR (G205.5+0.5). With a angular diameter of 3.6 degrees and an estimated distance around 1.6 kpc (but see \citealt{Zhao2018} for a different distance estimate) its has a physical diameter $\sim$100 pc. Despite its location nearly in the centre of the Galactic plane, it is believed to be expanding in a multi-phase ISM with a low density intercloud
medium ($n <$0.01 cm$^{-3}$; \citealt{Leahy1986}). 

Based on absorption features seen in background stars, the Monoceros Loop has an estimated expansion velocity between 50 and 70 km s$^{-1}$ \citep{Loz1972,Wallerstein1976}, not too dissimilar to that estimated for G107 from its optical spectra, with the exception of its northern and NW regions. The Monoceros remnant also exhibits numerous bright H$\alpha$ + [N~II] filaments that are  diffuse is appearance when imaged in [S~II] \citep{Kirshner1978}, again like G107. At a distance of 2 kpc, G107 would have a diameter similar to that of the Monoceros Loop of around 100 pc.

From the observed spatial separation of faint outer H$\alpha$ filaments ahead
of the bright, diffuse [O~III] filaments seen along the remnant's NW edge
(Figs.~\ref{O3_filament} \& \ref{O3_color}) we can constrain the density of the
local ISM in this region.  Figure~\ref{Raymond} shows two models of shock
emission for a 100 km s$^{-1}$ shock moving through an interstellar medium with
10\% neutral hydrogen at a distance of 1 kpc based on an updated version of the
model of \citet{Raymond1979}.  Here we modeled the H$\alpha$ + [N~II]
emission seen in the H$\alpha$ image shown in left panel of
Figure~\ref{O3_filament}. The top panel of Figure~\ref{Raymond} assumes a
preshock density $n_{\rm 0}$ =  0.1 $\rm cm^{-3}$ and a magnetic field of 1 $\mu$G,
while the bottom panel assumes $n_{\rm o}$ = 0.16 $\rm cm^{-3}$ and a field of 3
$\mu$G.  The asterisks and diamonds mark relative H$\alpha$ + [N~II] and
[O~III] flux measurements from our images, respectively, using a $3\arcsec
\times 3\arcsec$ box stepped every 3\arcsec\ in a perpendicular direction away
from the sharp H$\alpha$ filament marking the shock front location. The
horizontal axis scales inversely as the distance and preshock density, so the
density would be a factor of two lower at a distance of 2 kpc.  The geometry
was assumed to be a gently rippled sheet with an amplitude to wavelength ratio
of a few percent.

These simple models provide a surprisingly good match to the observed
[O~III] and H$\alpha$ + [N~II] emission to this perpendicular cut
across the filament.  The models predict a nearly unresolved H$\alpha$ +
[N~II] emission filament with a thickness of $\simeq 1''$ where neutral H
is excited and ionized just behind the shock. The H$\alpha$ drops when the
neutrals are ionized, and begins to increase as [N~II] emission increases
and the density increases as the gas cools.

In contrast, the [O~III] emission is very weak at the shock front,
increases in strengths as oxygen is ionized  to O$^{+2}$, then decreases as it
recombines.  This qualitatively agrees with what is seen along G107's NW limb,
where H$\alpha$ + [N~II] emission sharply decreases in strength
immediately behind the shock front, only to become stronger again some 20$''$
farther downstream. In addition, the [O~III] emission appears much broader
and distinctly diffuse, being some $\simeq 20'' - 25''$ in width, and slowly
rising and falling with increasing distance behind the shock front. We note
that this is similar to some filaments in the old Galactic Halo SNR G70.0-21.5,
where sharp H$\alpha$ is followed by more diffuse [O~III] emission
\citep{Raymond2020}. 

The model with $n_{\rm 0} = 0.1$ cm$^{-3}$ and B = 1 $\mu$G fits this particular cut
across the filament somewhat better than the model with 3 $\mu$G.  The magnetic
field at a few hundred pc in the Galactic halo is expected to be 3 to 4 $\mu$G
\citep{Sobey2019}.  At that field strength there is a significant magnetic
pressure behind the shock, while the low preshock density means a relatively
low thermal pressure.  Consequently, the magnetic field will inhibit
compression of the cooling gas and thus change the [O~III] and H$\alpha$
intensity profiles as functions of distance from the shock front, thereby
flattening the rise of H$\alpha$.  There is some variation among the measured
[O~III] and H$\alpha$+[N~II] profiles at different positions, and
the higher magnetic field might be preferred at at least one location.    

Based on the agreement of this model with our images,  we estimate $n_{\rm 0} = 0.13
\pm$ 30\% cm$^{-3}$ at 1 kpc, V$_{\rm shock} = 100 \pm 10$ km s$^{-1}$ for this NW
region.  The lower end of the density range would imply a magnetic field
somewhat lower than is typical for this height above the Galactic plane.  

The northern non-radiative Balmer emission filament requires a higher velocity, at least $\sim$150
km s$^{-1}$, to produce a higher post-shock temperature with no strong
[O~III] emission immediately behind the shock front. This implies a
density correspondingly lower  ($< 0.05$ cm$^{-3}$) and might also suggest a
higher neutral fraction to produce the bright Balmer emission. The reason why
this old remnant exhibits non-radiative filaments along a small section of its northern
limb and apparently some places elsewhere (see Fig.~\ref{Martins}) may be due to the remnant's location along the outer portion of the Galactic plane (Fig.~\ref{WIDE_MDW}) where it may have encountered small ISM regions of especially low density.

It is also
possible that the ram pressure is higher in this location, possibly because in
the past, the shock along G107's northern limb had moved through a lower
density medium than elsewhere in the remnant. Given the likely physical size of
the G107 remnant, large scale density variations leading to a range of ram
pressures would seem plausible in G107's later stages of evolution.

With the assumption that G107 is in its radiative phase, we can use the
formulae of  \citet{Chevalier1974} and \citet{Cioffi1988} for SNRs in the
radiative cooling phase.  Figure~\ref{CioffiModels} shows the
\citet{Cioffi1988} models for SNR radius and velocity as a function of time
using an explosion energy of $0.5\times10^{51}$ erg, at three different
preshock densities, $n_{\rm 0}$. The observed radius of the remnant is shown as
dashed lines on the left panel of the figure, where the minimum radius of
$\sim37$ pc corresponds to G107 being at a distance of 1.5 kpc, while the
maximum radius of $\sim 49$ pc if it were at 2 kpc. The shock velocity of the
[O~III] filament is shown as the dashed line in the right panel. 

We find agreement between the observed radius of the remnant at 2 kpc and the
model, when the density of matches that of the [O~III] filament, $n_{\rm 0} =
0.05$ cm$^{-3}$, at an age around $90 \times 10^{3}$ years old. However, these 1D models do
not match the filament's inferred shock velocity. This suggests the environment
through which G107 has and currently is expanding into is more complicated than
this simple uniform density model can account for. Given the large physical
size of the remnant, it is unlikely that it has expanded through a constant
density medium and this would affect both the remnant's radius and shock
velocity.  Using these same models for the somewhat larger G70.0-21.5 remnant,
which has a higher estimated preshock density compared to G107, there was a
similar over-prediction of the shock velocity for a given radius
\citep{Raymond2020}. 

Given that the velocity estimate of 100 km s$^{-1}$ for G107's [O~III]
bright NW region may represent the remnant's higher expansion velocity region
and hence among its lowest ambient preshock density, then higher preshock
densities and hence lower shock velocities are then likely present in most
other  optically bright regions of the remnant. This would appear especially
true along most of G107's western limb where the shock velocity appears to be
less than 70 km s$^{-1}$  due to the lack of [O~III] emission
\citep{Raymond1979,Shull1979}. Consequently, such large and denser ISM regions
may be more representative of the remnant's environment and evolutionary age.

In a remnant's cooling phase, its age, $t_{\rm cool}$
can be estimated  from \citet{Falle1981}
\begin{equation}
    t_{\rm cool} = 2.7 \times 10^{4} ~ E_{51}^{0.24} ~ n_{\rm 0}^{-0.52} ~ ~ {\rm yr}
\end{equation}
where E$_{51}$ is energy in units of $10^{51}$ erg and $n_{\rm 0}$ in the preshock density in cm$^{-3}$.
Adopting $E_{51}$ = 0.5 and a preshock densities $\simeq0.050 - 0.067$ cm$^{-3}$ based on the [O~III] filament's ambient density at distances of 1.5 and 2 kpc, the above expression gives an estimate range of G107's age,   
$t_{\rm cool}$, of $\sim90 - 110\times 10^{3}$ yr. 

Although other distances to G107 than $1.5 - 2.0$ kpc are possible, as noted
above the spectra of its NW and western filaments together with the observed
H$\alpha$ and [O~III] emission separation place some constraints on G107's
distance in regard to the remnant's shock velocities and preshock densities in
the regions we examined. Further investigations into the few filaments present
along G107's eastern  and northeastern limbs would help constrain its expansion
in those directions and hence better determine its overall expansion and shock
properties.

\section{Conclusions}

We report the discovery of a new large supernova remnant with an extensive
optical emission structure. Wide-field H$\alpha$ images show a $\simeq2.8\degr$
diameter supernova remnant centred near $l$ = 107.0, $b$ = +9.0. Higher
resolution H$\alpha$ and [O~III] 5007 \AA \ images reveal an extensive
complex set of thin and overlapping filaments along the remnant's northern,
western, and southwestern limbs, with few filaments bright in [O~III]
emission.  

Optical spectra show both Balmer dominated, non-radiative filaments and
brighter radiative filaments with [S~II]/H$\alpha$ ratios consistent with
shock emission. Filament emission lines suggest a range of shock velocities,
around $< 70$ km s$^{-1}$ along its western limb, with higher shock velocities
$\simeq100 - 150$ km s$^{-1}$ along its northern and northwestern boundary. We
estimate the remnant to be physically large with a diameter $\sim75 - 100$ pc, located
at a distance of $1.5 - 2$ kpc and lying some $250 - 300$ pc above the Galactic
plane. 

Follow-up moderate to high-dispersion spectroscopy on bright stars lying behind the remnant could be used to determine both the remnant's distance and general expansion velocities like that done for some other Galactic SNRs (e.g., \citealt{Wallerstein1976,Jenkins1984,Fesen2018,Ritchey2020}). Associated radio and X-ray emissions could also be searched for especially near its brighter optical emissions. The finding of such a large and optically bright remnant might suggest other, high Galactic latitude SNRs still await discovery.

\bigskip

\section*{acknowledgements}

We thank Justin Rupert, Eric Galayda
and the MDM staff for making these observations possible, and Sakib Rasool for helpful communications.
K.E.W.\ acknowledges support from Dartmouth's Guarini School of Graduate and
Advanced Studies, and the Chandra X-ray Center under CXC grant GO7-18050X.
This work is part of R.A.F's Archangel III Research Program at Dartmouth.

\section*{Data Availability}
The H$\alpha$ MDW images presented here are available through the MDW Hydrogen-Alpha Sky Survey (www.mdwskysurvey.org/). The remaining wide-field images 
are availible through contact with the co-authors at https://www.cxielo.ch/ and at http://www.sternwarte-baerenstein.de/.
The optical images and spectra obtained at MDM Observatory presented here are available upon request from the corresponding author. 




\bibliographystyle{mnras}
\bibliography{ref2}

\bsp	
\label{lastpage}
\end{document}